\documentclass[12pt]{article}
\usepackage{amsmath}
\usepackage{graphicx}
\usepackage{geometry}
\usepackage[applemac]{inputenc}
\usepackage{natbib}

\ifx\pdfoutput\@undefined\usepackage[usenames,dvips]{color}
\else\usepackage[usenames,dvipsnames]{color}
\IfFileExists{pdfcolmk.sty}{\usepackage{pdfcolmk}}{} 
\fi
\usepackage[plainpages=false,pdfpagelabels,pagebackref=false,naturalnames=true,hyperindex=true,pdftitle={***},pdfauthor={Carlos Gershenson}]{hyperref}
\hypersetup{colorlinks=true,
urlcolor=Cerulean,linkcolor=BrickRed,citecolor=RoyalBlue,
  pdfpagemode=None,
  pdfstartview=FitH}
\usepackage[all]{hypcap}

\begin{document}

\title{Guiding the Self-organization of Cyber-Physical Systems} 

\author{Carlos Gershenson$^{1,2,3}$ \\
$^{1}$ Departamento de Ciencias de la Computaci\'on\\
Instituto de Investigaciones en Matem\'aticas Aplicadas y en Sistemas \\
Universidad Nacional Aut\'onoma de M\'exico\\
A.P. 20-726, 01000 M\'exico CDMX M\'exico\\
\href{mailto:cgg@unam.mx}{cgg@unam.mx} \
\url{http://turing.iimas.unam.mx/~cgg} \\
$^{2}$ Centro de Ciencias de la Complejidad \\
Universidad Nacional Aut\'onoma de M\'exico\\
$^{3}$ ITMO University, Russian Federation.}
\maketitle

\begin{abstract}

Self-organization offers a promising approach for designing adaptive systems. Given the inherent complexity of most cyber-physical systems, adaptivity is desired, as predictability is limited. Here I summarize different concepts and approaches that can facilitate self-organization in cyber-physical systems, and thus be exploited for design. Then I mention real-world examples of systems where self-organization has managed to provide solutions that outperform classical approaches, in particular related to urban mobility. Finally, I identify when a centralized, distributed, or self-organizing control is more appropriate.
\end{abstract}

\section{Introduction}

We are submerged in \textbf{complexity}. And this complexity is increasing. But what is complexity? There are dozens of definitions and measures in the literature \citep{lloyd2001measures,GershensonHeylighen2005}, but not a definite one. Well, life is not properly defined either, and it is not a hindrance for biology. Still, to have an idea of what we refer to, let us go to its etymological root. Complexity comes from the Latin \emph{plexus}, which means entwined. In other words, something complex is difficult to separate. This is because the interactions among its components are relevant \citep{Gershenson:2011e}. Relevant because they co-determine the future of the system. Thus, if we do not consider such interactions, but study components in isolation, we will not be able to understand the system properly. Also, interactions can generate novel information, not present in initial nor boundary conditions. This novel information limits predictability \citep{Gershenson2013Facing-Complexi} and is the source of computational irreducibility \citep{Wolfram:2002}, \emph{i.e.} there is no shortcut to know the future: one must go through all intermediate steps, because the information produced in the process is required to reach/compute the future.

A recent collaborative effort produced this definition: ``Complexity science, also called complex systems science, studies how a large collection of components --- locally interacting with each other at small scales --- can spontaneously self-organize to exhibit non-trivial global structures and behaviors at larger scales, often without external intervention, central authorities or leaders. The properties of the collection may not be understood or predicted from the full knowledge of its constituents alone. Such a collection is called a complex system and it requires new mathematical frameworks and scientific methodologies for its investigation.'' \citep{ComplexityExplained}

One of the core concepts explained in \citet{ComplexityExplained} is \textbf{self-organization}: ``Interactions between components of a complex system may produce a global pattern or behavior. This is often described as self-organization, as there is no central or external controller. Rather, the ``control'' of a self-organizing system is distributed across components and integrated through their interactions. Self-organization may produce physical/functional structures like crystalline patterns of materials and morphologies of living organisms, or dynamic/informational behaviors like shoaling behaviors of fish and electrical pulses propagating in animal muscles. As the system becomes more organized by this process, new interaction patterns may emerge over time, potentially leading to the production of greater complexity.'' Common examples of self-organizing systems \citep{CamazineEtAl2003} include flocks of birds, schools of fishes, insect swarms, herds, crowds, and other collective phenomena \citep{Vicsek2012}, although self-organization is not restricted to living systems \citep{NicolisPrigogine1977,Haken1988,GershensonHeylighen2003a}.

There are many cases where self-organization has been used as an approach in \textbf{engineering} \citep{EngineeringSOS2004,DeWolfEtAl2005,ZambonelliRana2005}. In this case, we can \emph{describe} a system as self-organizing when elements interact to achieve dynamically a global function or behavior \citep{GershensonDCSOS}. In other words, instead of designing directly a solution, one regulates the potential interactions among elements. This is useful in \emph{non-stationary} problems: when the situation changes, then the system adapts by itself. Since interactions in complex systems produce novel information, it is common that this information will change a complex problem. Not only its state, but also its state space. Thus, self-organization can be useful to face complexity by providing general adaptation mechanisms. Several methodologies using self-organization have been proposed (see \citet{Frei:2011} for an overview), although the approach has not been widely applied.

In a parallel effort, \textbf{guided self-organization} attempts to combine seemingly opposed processes: design to define and regulate the properties and behavior of a system, and self-organization that implies certain autonomy and adaptability \citep{Prokopenko:2009,GSOInception2014} Guided self-organization can be understood as ``the steering of the self-organizing dynamics of a system towards a desired configuration'' \citep{Gershenson:2010}.

In this paper, I complie concepts and approaches useful for designing self-organizing systems in the physical realm. I illustrate these with case studies from urban mobility before concluding.

\section{Concepts}

Several concepts are useful to design and guide self-organizing systems. In this section, a non-exhaustive list is presented.

\subsection{Adaptation}

Adaptation can be defined as a change in an agent or system as a response to a state of its environment that will help the agent or system to fulfill its goals \citep{GershensonDCSOS}. Living systems naturally adapt to changes in their environment, and artificial systems can benefit from exhibiting adaptation \citep{Holland1975,steels1995artificial,Bedau2013IntroductionLT}.

If problems are \textbf{stationary}, \emph{i.e.} do not change, then it is worthwhile attempting to predict the future of a system to control it. However, for \textbf{non-stationary} problems, predictability by definition is limited. Novel information generated by interactions in complex systems can lead to non-stationarity. In this case, adaptation is desirable to complement the unpredictable aspects of a problem \citep{Gershenson2013Facing-Complexi}. 

For example, city traffic is changing constantly: every time a red light switches to green, the number of vehicles waiting changes. Thus, the timing of the traffic lights should also change. Traditional adaptive traffic light control methods (\emph{e.g.} Sydney, Dublin, Singapore) use sensors to shift phases depending on recent average demands. This is usually better than not having adaptation, where the best possible option would be to take average measurements, set fixed phases, and perhaps change the programs a few times per day. However, if traffic lights can adapt at the same timescale as the traffic demand does, \emph{i.e.} every cycle, then the performance would be much improved \citep{goel2017self}.

\subsection{Robustness}

A system is \textbf{robust} if it continues to function in the face of perturbations \citep{Wagner2005}. As with adaptation, robustness is prevalent in living systems and desirable in artificial ones \citep{Jen2005}.

Robustness and adaptability are complementary: a system has to be robust enough to survive while it adapts, and adaptation can favor robustness.

For example, the Internet is quite robust. The TCP/IP protocol was designed to resist nuclear warfare. At the content level, self-organization has led to a scale-free topology \citep{BARABASI200069}, which is also robust to random failures (although fragile to directed attacks).

\subsection{Antifragility}

A fragile system is damaged by perturbations. A robust system is unaffected by perturbations. An \textbf{antifragile} system \emph{benefits} from perturbations \citep{Taleb2012}.

For example, the immune system is antifragile. Children who grow up in extremely sanitized conditions are not exposed to pathogens (perturbations), so their immune systems do not develop, leading to stronger infections in adulthood. Certainly, children should not be infected intentionally, but being exposed to a ``normal'' amount of pathogens and falling ill now and then helps train the immune system.

We have recently proposed a measure of antifragility \citep{Pineda2019}, which captures the idea of being positive when perturbations improve the performance of a system, is negative when perturbations decrease the performance (fragility), and is zero when perturbations do not affect the performance (robustness). An important aspect is that there is no ``optimal'' antifragility independent of an environment. A system should be as antifragile as its environment varies (this is related with requisite variety, discussed in Section~\ref{sec:approaches}). 

\subsection{Mediators}

Interactions can be classified as positive, neutral, or negative, depending on the effect they have on the goals of a system \citep{GershensonDCSOS,Gershenson:2010a}

A \textbf{mediator} arbitrates among the elements of a system, to minimize conflict, interferences and frictions (negative interactions); and to maximize cooperation and synergy (positive interactions)\citep{Michod2003,Heylighen:2006,GershensonDCSOS}.

For example, traffic rules aim at reducing conflict in urban mobility. Without these rules, we would need to decide constantly on which side of the streets to drive, how to give way, make turns, etc. Even when rules and norms vary from country to country, and in some cases from city to city, when everybody follows the same set of rules (mediators), conflicts tend to be reduced. If they were not, then the rules should be changed.

Designing mediators can be useful for regulating systems where the elements cannot be modified. Still, mediators can change the interactions between elements, leading to different systemic behavior and properties (See case study in Section~\ref{sec:crowdcontrol}).

\subsection{Slower-is-faster effect}

Probably this effect was first described less than twenty years ago \citep{PhysRevLett.84.1240,Helbing:2000} while modeling crowd dynamics. If people trying to evacuate a room are panicked (trying to exit faster), then they create friction which leads to a ``turbulent'' flow that is slower than if people exit calmly, thus with a ``laminar'' flow. The same effect has been studied in vehicular traffic, logistics, public transport, social dynamics, ecological systems, and adaptive processes \citep{GershensonHelbing2015}. 

In general, the slower-is-faster effect occurs when a system performs worse as its components try to do better. This implies that a balance between doing ``too few'' and doing ``too much'' is necessary. However, in many cases this balance is dynamic. For example, the optimal speed for highway traffic (that maximizes flow) depends on the vehicular density. For this reason, systems that present a slower-is-faster effect, require constant adaptation, that can be achieved through self-organization.

The slower-is-faster effect may refer to any variable, not only speed. For example, growth or profits are not necessarily maximized in the long term with a short-term maximization strategy. Managing natural resources, such as fisheries, requires this understanding: if all resources are depleted, then in the near future there will be no profits. Maximizing profits requires a careful balance between short-term action and long-term planning. As with the case of highway traffic, usually this balance is non-stationary.

\subsection{Heterogeneity}

Most of our models of complex systems are homogeneous: all components have the same properties. This simplification is useful when we face computational limitations. However, increasing processing power and data availability have allowed us to make more realistic models, where different elements of a system have varying properties.

Perhaps the most studied heterogeneity in complex systems is the one of network topologies \citep{AlbertBarabasi2002,NewmanEtAl2006,GershensonProkopenko:2011,Barabasi2016}. Different organizations of the same elements can lead to radically different functionalities. A classical example is different arrangements (allotropes) of carbon atoms, which can lead to charcoal, diamond, graphite, graphene, nanotubes, buckyballs, etc. The components are the same, but changing their organization leads to radically different properties of these materials. Many networks are heterogeneous, with few elements having lots of connections and many elements having few connections.

More recently, temporal heterogeneity has been also studied \citep{Cocho2015,10.3389/fphy.2018.00045}. In a similar way, few elements change slower than most elements that change faster. This heterogeneity seems to lead to a balance where slow elements are robust and fast elements are adaptable. In homogeneous systems, this balance is achieved only in phase transitions, which can be characterized as ``critical'' \citep{Balleza:2008}. However, heterogeneity seems to expand the balance beyond criticality, making it easier to search an unknown parameter space, simply because different components diversify any search procedure.

\section{Approaches}
\label{sec:approaches}

How to implement the properties related to self-organization in cyber-physical systems?
The concept of self-organizing systems originated within cybernetics \citep{Ashby1947sos,vonFoerster1960,Ashby1962}, where useful approaches were already developed.

Ashby not only coined the term ``self-organizing system'', but he also proposed the law of \textbf{requisite variety} \citep{Ashby1956,HeylighenJoslyn2001,BarYam2004,Gershenson2015Requisite-Varie}. Variety can be understood as the possible number of states that a system can have. This law states that a controller must have at least as much variety as the system it is trying to control. For example, if we want a robot at a manufacturing plant to deal with seven different types of boxes, then it should be able to distinguish and make the appropriate decisions to handle each type of box. A common problem is that complexity explodes variety and vice versa. Therefore, traditional approaches become limited. To handle the variety of a system, we can either reduce its variety (using mediators), or increase the variety of the controller, but then the latter will imply an increase in the complexity of the controller as well.

There is an interesting relationship between variety and \textbf{heterogeneity}. Heterogenous systems by definition have more variety, so in principle they should be able to control more situations than similar homogeneous systems. However, they might be less robust and more complicated to design and understand. For example, ``if there is a system of ten agents each able to solve ten tasks, a homogeneous system will be able to solve ten tasks robustly (if we do not consider combinations as new tasks). A fully heterogeneous system would be able to solve a hundred tasks, but it would be fragile if one agent failed.'' \citep[p. 53]{GershensonDCSOS}. Thus, a balance between homogeneity and heterogeneity should also give us a balance between \textbf{robustness} and \textbf{adaptability} \citep{Langton1990,Kauffman1993}.

We can consider computers as telescopes of \textbf{complexity} \citep{Pagels1989}. In other words, without computers, our cognitive abilities are limited to studying models considering only two or three variables. To explore models with thousands or millions of variables, \textbf{computer simulations} are necessary \citep{GershensonDCSOS} because of computational irreducibility \citep{Wolfram:2002}. Complexity implies that new information is generated by interactions, so there is no ``shortcut'' to the future and all intermediate steps are necessary. This limits inherently the predictability of systems \citep{Gershenson2013Facing-Complexi}.

\textbf{Agent-based modeling} \citep{Bonabeau2002ABM,epstein2006,Wilensky2015} has been a useful approach to describe complex systems. Considering elements as agents, with states, goals, and rules allows us to study how changes at one scale lead to effects at another scale. The effects can go in both directions: changes in agents leading to changes in the system and vice versa.

Another approach that is becoming more and more popular as data availability and computing power increase is \textbf{network science} \citep{Newman:2003,NewmanEtAl2006,Barabasi2016}. Networks have the benefit of being able to represent naturally elements (nodes) and interactions (links). The relationship between the structure and function of networks has been an intense area of study, where self-organization can play a relevant role \citep{Gershenson:2010}.

Ethology --- the study of animal behavior --- has been taken as an inspiration to build \textbf{adaptive} systems \citep{Beer1990,Maes1994,steels1995artificial} and to study complex artificial systems
\citep{Rahwan2019}. Animals have evolved to survive in complex environments, so adaptive strategies and self-organizing mechanisms found in nature have been used in cyber-physical systems. In this sense, living technology \citep{Bedau:2009,GershensonALife2018} takes the advantageous properties of living systems and applies them in socio-technical systems, from protocells \citep{Protocells2008} to cities \citep{Gershenson:2013}.

The \textbf{robustness} and \textbf{antifragility} of systems can be promoted through different mechanisms \citep{Gershenson:2010}, such as redundancy (having several copies of the same element), degeneracy (having different elements perform the same function), modularity (short-range links stronger than long-range ones), and scale-free-like (heterogeneous) topologies (few elements with several links, several elements with few links) .

\section{Case studies}

In this section, I illustrate the previous concepts and approaches with case studies we have worked with in recent years, related to urban mobility. 

\subsection{Crowd control}
\label{sec:crowdcontrol}

More than a hundred million people use the hundred busiest metro systems in the world every day, a number that is growing fast as the urban population is increasing and cities develop. In the Mexico City Metro and other systems, people would normally push each other, not letting passengers exit trains, collapsing the systems. How to regulate passenger behavior, when a selfish approach might seem to bring individual benefit but lead to collective inefficiency?
One can think of different \textbf{mediators}, but they can be costly to try in real systems. To explore alternatives, we first used crowd simulations \citep{Helbing:2000} and then implemented a pilot study in the Balderas station of the Mexico City Metro on December, 2016 \citep{10.1371/journal.pone.0190100}. The pilot was a success and it has since been extended to several other busy stations.

\begin{figure}[htbp]
\begin{center}
\includegraphics[width=0.66\textwidth]{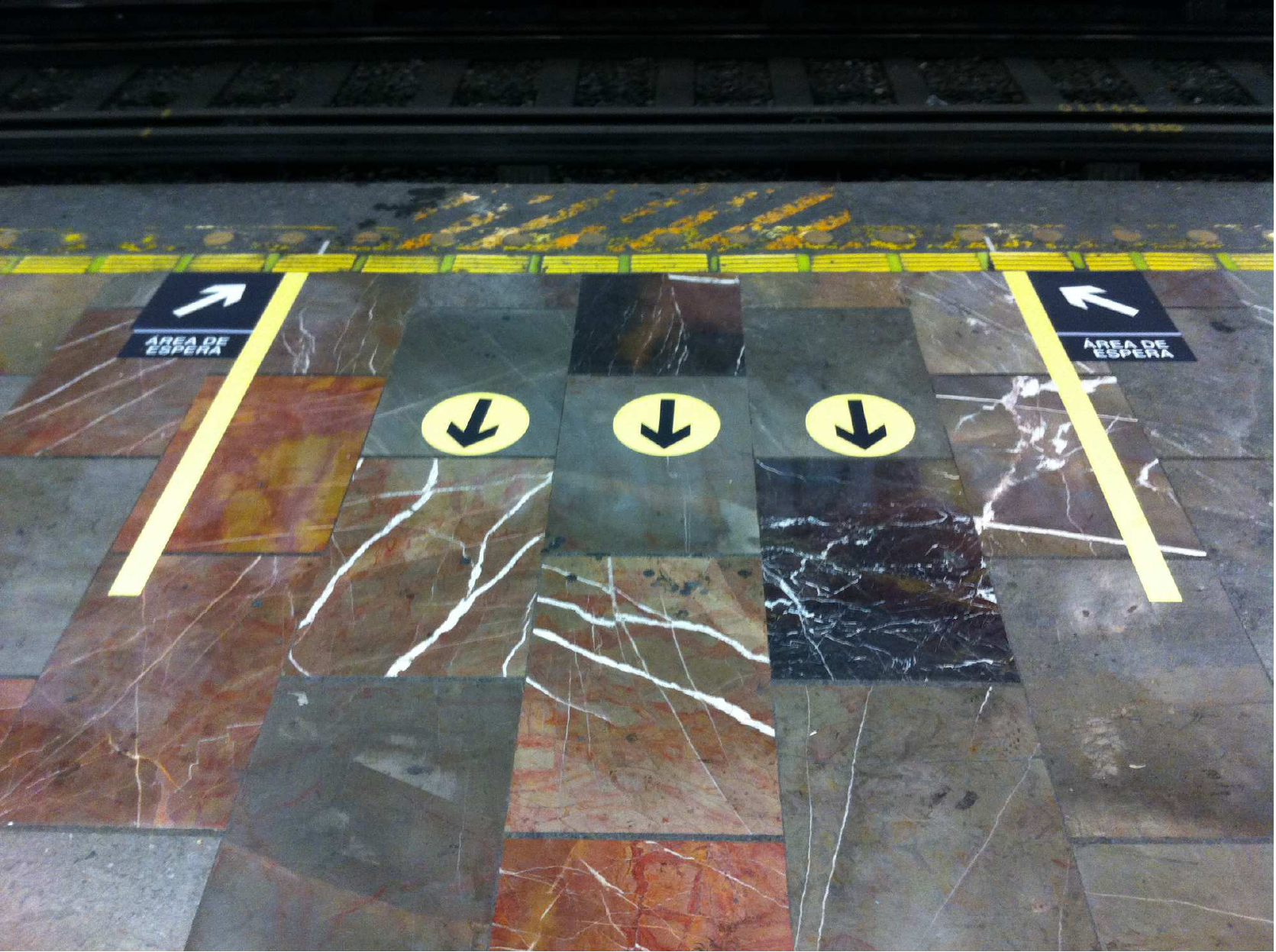}
\caption{Signs installed to mediate passenger boarding and descent in Mexico City Metro \citep{10.1371/journal.pone.0190100}.}
\label{fig:metro-signs}
\end{center}
\end{figure}

\begin{figure}[htbp]
\begin{center}
\includegraphics[width=0.66\textwidth]{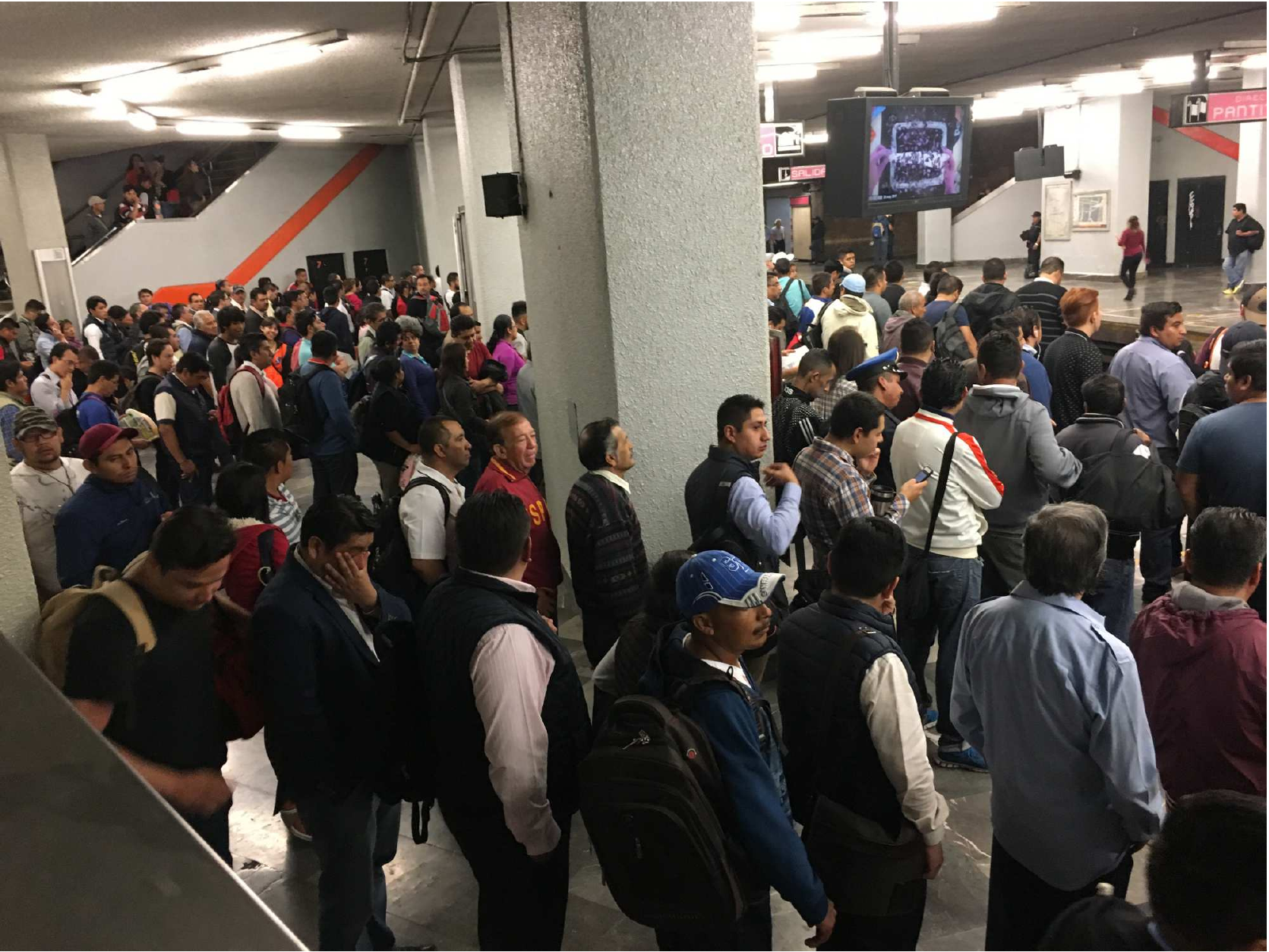}
\caption{Passengers queuing waiting for a train in Mexico City Metro during rush hour, San L\'azaro metro station \citep{10.1371/journal.pone.0190100}.}
\label{fig:metro-queues}
\end{center}
\end{figure}

The intervention consisted of ``simple'' signs that indicate passengers roughly where the train doors will be, asking them to leave free space for exiting passengers, as shown in Figure~\ref{fig:metro-signs}. What we did not expect nor suggest was that people would queue (Figure~\ref{fig:metro-queues}), and that these queues could even go upstairs as people respected them.

This intervention managed to change the behavior of the passengers and thus the crowd, without changing the elements of the system (where could we get different ``educated'' passengers from?). The signs mediated \emph{interactions} between people. This is an example of a \textbf{passive} control, where interactions are regulated ``simply'' providing useful information.

\subsection{Traffic light coordination}

The coordination of traffic lights is an EXP-complete problem, meaning that in theory it takes exponentially more time to find a solution as more intersections are added to a street network. Also, the precise number of vehicles changes every cycle, so in practice the problem changes faster than it can be optimized. An \textbf{active} controller should \textbf{adapt} as fast as the controlled changes (\textbf{requisite temporal variety}), and for that sensors are required to provide relevant information to the controller.

With this in mind, we have proposed self-organizing algorithms that can coordinate traffic flows and adapt to constant changes in the demand as fast as it changes \citep{Gershenson2005,Zapotecatl2017}, achieving close-to-optimal performance \citep{GershensonRosenblueth:2011}. The main idea behind the algorithms is that streets with a higher demand get a preference. Thus, busier directions will wait less for a green light. This increases the probability that vehicles will aggregate behind red lights with few cars, leading to the formation of platoons. As platoons reach a certain size, they can request a green light before they even reach an intersection, so vehicles do not need to stop, unless there are other vehicles or pedestrians crossing. Platoons are easier to coordinate than individual vehicles, as they leave spaces between them that other platoons can use without interference. When densities are high, the preference is given to the street that has more space after the intersection, preventing gridlocks.

It is difficult to compare the performance of self-organizing traffic lights, as there are no benchmarks in traffic light coordination. However, they are close to optimal. Also, we have found that self-organizing traffic lights would improve traffic more than if all vehicles were autonomous but with traditional traffic lights. Nevertheless, autonomous vehicles and self-organizing traffic lights are even better.

By distributing control locally (we have made simulations with up to ten thousand intersections achieving efficient or optimal coordination), the \textbf{requisite variety} of the traffic light coordination can be tackled as conditions change, while the formation of platoons self-organizes the traffic flows and assists the coordination of intersection controllers at the city scale. In this way, the traffic lights are \textbf{mediators} of vehicles, but the vehicles are also \textbf{mediators} of traffic lights.

\subsection{Public transport regulation}

In theory, passengers in public transport are served optimally when vehicle headway --- the time between arrivals at a station --- is equal. However, as we have shown, an equal headway configuration is unstable by nature \citep{GershensonPineda2009}, as delays become amplified by positive feedbacks. Thus, many approaches are taken by transportation engineers to prevent the ``equal headway instability'', also known as the ``bus bunching problem''. 

To keep equal headways, all vehicles --- trains, trams, buses --- must wait the same time at each station. This time can vary from station to station, but it must be fixed or some vehicles will go faster than others, leading to unequal headways and potentially to the collapse of the system.  Since the precise number of passengers varies each time a vehicle reaches a station, and thus the required waiting time, then either vehicles will require a margin and be idle, or they will depart before servicing all passengers when these are more than expected.

We proposed a self-organizing algorithm inspired by ant colony communication \citep{Gershenson:2011a,10.1371/journal.pone.0190100}, where each vehicle ``simply'' tries to keep equal distance to the vehicles in front and behind, but is flexible enough to serve passengers at stations and at the same time prevent idling. Equal headways are not maintained, but the system does not collapse. Rather, its performance is even better than the case with equal headways, \emph{i.e.} it is supraoptimal. This is because of the \textbf{slower-is-faster effect}: It is true that passengers minimize their waiting time at stations with equal headways (what theory says). But their total travel time is not independent of the equal headways, so idling will increase their total travel time. With the self-organizing algorithm, passengers wait more at stations, but once they board a vehicle, they will reach their destination faster, as there is no idling. Again, \textbf{adaptation} takes place at the scales at which the system changes.

\section{Discussion}

We cannot reduce the complexity of several systems we have to deal with. Novel information produced by interactions leads to changes, making problems non-stationary. Self-organization has been used in a broad variety of cyber-physical systems. It allows systems to adapt at the scales at which the problem they are solving changes in a robust fashion. In addition to the case studies mentioned in the previous section, dynamic road pricing in Singapore and variable parking cost in San Francisco are examples of self-organization being used to regulate urban mobility. We can see that the same principles apply in other cyber-physical systems, from telecommunications \citep{Amoretti2015Measuring-the-c} to organizations \citep{GershensonSOBs}.

A relevant step towards adopting self-organizing controllers is to give up the desire to control completely our systems. As complexity limits our predictability, systems require certain autonomy to make the ``right decisions''. Even if we use traditional approaches, we do not have full control of our systems, as they are constantly entering unexpected situations. We would like to be able to be sure that our systems will never fail, but they will. We can have formal proofs but these are also limited, since they assume idealized/closed/predefined situations. Self-organizing systems can do the same as traditional engineered systems and more, as they can deal with more realistic/open/variable situations. We just have to try and see, constantly adapting \citep{GershensonDCSOS}. Even if a solution already worked, it does not assure that it will continue working (as conditions change) or that it can be applied in the same way in a different context.

The best solution depends on the context/environment/problem. In some cases, centralized control will be good, in others distributed is more appropriate, in yet others self-organizing. As shown in Table~\ref{tab:controls}, \textbf{centralized} control is appropriate when causality should be top-down. Because of the law of requisite variety, systems with a high variety/complexity will require a controller with a high variety/complexity, so the centralized approach becomes less viable. \textbf{Distributed} control can deal with a greater complexity, but it is still limited, because the integration of the distributed solutions is not necessarily trivial. This limits distributed control to homogeneous systems: since information flow across the system is limited, the local solutions assume that each local problem is similar.
\textbf{Self-organizing} control can deal with top-down and bottom-up causality (multiscale), as components can interact in a distributed fashion to change system properties (bottom-up), but then the system properties can mediate (top-down) to regulate the behavior of components. Self-organization can be scalable, adaptive, robust, and can deal with a high complexity and homogenous or heterogeneous problems. It is not that one approach is better than others, but they are more appropriate for different problems. Centralized control is easier to implement and understand, but is useful for low complexity/variety problems. Distributed control can deal with a greater complexity, but only for heterogeneous systems. Self-organizing systems might be more difficult to design and test, but they can handle greater complexity/variety/diversity.

\begin{table}[htp]
\caption{Different control approaches are more appropriate for different causalities, complexities, and diversities.}
\begin{center}
\begin{tabular}{|c|c|c|c|}
\hline
\textbf{Control} & \textbf{Causality} & \textbf{Complexity} & \textbf{Diversity}\\
\hline
Centralized & top-down & low & homogeneous or heterogeneous \\
\hline
Distributed & bottom-up & high & homogeneous \\
\hline
Self-organizing & multiscale & high & homogeneous or heterogeneous \\
\hline
\end{tabular}
\end{center}
\label{tab:controls}
\end{table}%

As the complexity of our cyber-physical systems increases, and also our understanding of it, we will see more self-organizing approaches. Perhaps names will differ, but the concepts presented here are required to control cyber-physical systems by guiding their self-organization.

\bibliographystyle{cgg}

\bibliography{refs,carlos}

\end{document}